\documentclass[sigconf]{acmart}

\usepackage{booktabs} 
\usepackage{bbm}
\usepackage{algorithm}
\usepackage{algpseudocode}
\usepackage{enumitem}
\usepackage{multirow}

\setcopyright{rightsretained}

\acmConference[recsysXfashion'19]
{Workshop on Recommender Systems in Fashion, 13th ACM Conference on Recommender Systems}
{September 20, 2019}
{Copenhagen, Denmark}
\acmYear{2019}
\copyrightyear{2019}

\acmArticle{}
\acmPrice{}
\acmDOI{}
\acmISBN{}

\begin{document}
    \title[]{ Session-based Complementary Fashion Recommendations}

\author{Jui-Chieh Wu}
\affiliation{%
\institution{Zalando SE}
\streetaddress{Muehlenstrasse 25}
\city{Berlin}
\state{Berlin}
\postcode{10243}
}
\email{jui-chieh.wu@zalando.de}

\author{Jos\'{e} Antonio S\'{a}nchez Rodr\'{i}guez}
\affiliation{%
\institution{Zalando SE}
\streetaddress{Muehlenstrasse 25}
\city{Berlin}
\state{Berlin}
\postcode{10243}
}
\email{jose.antonio.sanchez.rodriguez@zalando.de}

\author{Humberto Jes\'{u}s Corona Pamp\'{i}n}
\affiliation{%
\institution{Zalando SE}
\streetaddress{Muehlenstrasse 25}
\city{Berlin}
\state{Berlin}
\postcode{10243}
}
\email{humberto.corona@zalando.de}

\renewcommand{\shortauthors}{Wu et al.}

    \begin{abstract}

    In modern fashion e-commerce platforms, where customers can browse thousands to millions of products,
    recommender systems are useful tools to navigate and narrow down the vast assortment.
    In this scenario, complementary recommendations serve the user need to find items that
    can be worn together.
    In this paper, we present a personalized, session-based complementary item recommendation algorithm, \textit{ZSF-c},
    tailored for the fashion usecase.
    We propose a sampling strategy adopted to build the training set, which is useful when existing user
    interaction data cannot be directly used due to poor quality or availability.
    Our proposed approach shows significant improvements in terms of accuracy compared to the collaborative filtering
    approach, serving complementary item recommendations to our customers at the time of the experiments (\textit{CF-c}).
    The results show an offline relative uplift of +8.2\% in Orders Recall@5, as well as a significant +3.24\% increase
    in the number of purchased products
    measured in an online A/B test carried out in a fashion e-commerce platform with 28 million active customers.

\end{abstract}

    %
%

\ccsdesc[500]{Retrieval tasks and goals~Recommender systems}
\ccsdesc[500]{Computing methodologies~Neural networks}

\keywords{Session-based recommendation, Fashion, Complementary item recommendation, Neural networks}

    \maketitle

\section{Introduction}

\begin{figure}[ht]
    \centering
    \includegraphics[width=0.35\textwidth]{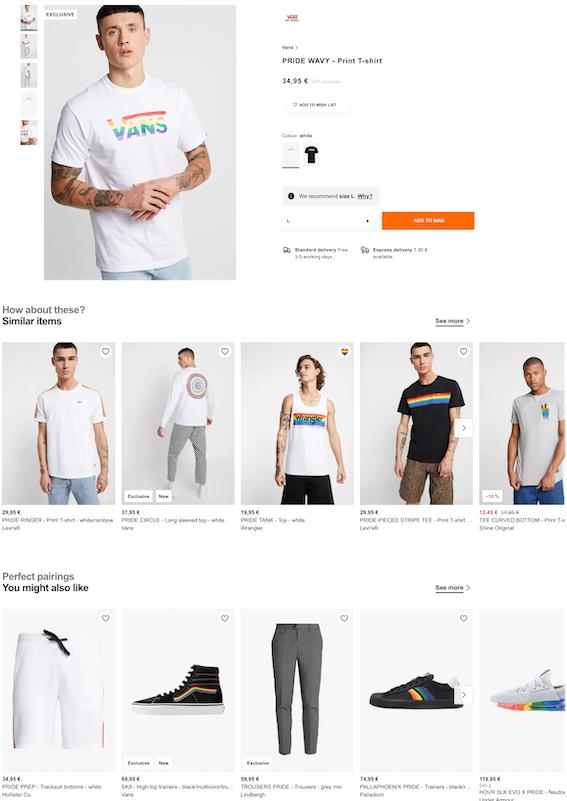}
    \caption{Example of a Product Display Page showing different recommendations.
    The complementary item recommendations are shown under the heading \textit{"Perfect pairings: You might also like"}.
    These type of recommendations allow customers to continue
    the fashion discovery journey by finding items that can be worn together. }
    \label{fig:screenshot}
\end{figure}

Recommender systems provide a solution for different user needs, such as; finding items similar to a given one,
building outfits \cite{gomes2017boosting}, or discovering what to wear for a special occasion.
In Zalando, a fashion e-commerce platform in  Europe,
one of the most-used recommendation tools is the complementary item recommendations.
These recommendations are displayed in the Product Display Page (PDP), which is one
of the pages with more traffic in the platform. An example of a PDP is shown in Figure \ref{fig:screenshot}.

The complementary item recommendation in production at the time of the experimentation was
based on a collaborative filtering approach \cite{ aiolli2013preliminary}, denoted as \textit{CF-c}.
It works by finding the most similar items to a given one (based on cosine similarity) that satisfy the definition of complementary.
However, as often the amount of recommended items that satisfy the definition of complementary is not enough, we
expand the recommendations by finding items that are similar to these original recommendations, they must also satisfy the
definition of complementary.
While this approach, together with some basic business rules, generally perform well in terms of click through rate
(CTR), it has several flaws that led to poor customer experience, which has been highlighted in explicit feedback
from customers.
Moreover, it is not serving the purpose of helping customers find items that are complementary to a product they have
expressed interest on, as reflected in the performance of orders attributed to this type of recommendations.

Recent studies on session-based Recommendations have
shown significant improvements compared to collaborative filtering approaches in several
datasets \cite{HidasiKBT15, Hidasi:2018:RNN:3269206.3271761, Li:2017:NAS:3132847.3132926, Liu:2018:SSA:3219819.3219950}.
Moreover, they have also proven succesfully in previous internal efforts, to tackle similar recommendation problems.
Thus, we use the approach proposed in \textit{STAMP} \cite{Liu:2018:SSA:3219819.3219950},
and introduce several improvements to optimize the model performance for the particular task of complementary fashion
item recommendations.
For example; adding order events, and category and image embeddings.
We refer to this proposed approach as \textit{ZSF-c}, \textit{Zalando STAMP Fashion Complementary Recommendations}.

\vspace{0.2cm}

The rest of the paper is organised as follows.
First, Section \ref{section:related_work} describes the related work.
Then, Section \ref{section:problem-statement} formalizes the problem.
Section \ref{section:proposed-method} describes the proposed approach and dataset.
Section \ref{section:experiments} introduces the preliminary offline and online results.
Finally, Section \ref{section:conclusions} presents conclusions and future work.

\section{Related Work}
\label{section:related_work}

As complementary items recommendations become more and more important to improve the customer journey in
e-commerce platforms, there has been an increase on the research focus in the topic.
In \cite{DBLP:journals/corr/abs-1809-09621}, Trofimov proposes \textit{BB2vec}, a scalable and extendable
complementary item recommendations algorithm. The approach includes both browsing and purchase data to
alleviate the cold start problem, when compared to approaches that only exploit purchase data.

Zahang et. al. \cite{Zhang:2018:QNC:3240323.3240368} propose \textit{Encore}, a neural complementary recommender
that learns complementary item relationships and user preferences jointly.
This approach, which is able to combine both stylistic and functional facets of complementary items across
categories, reports an improvement of 15.5 \% accuracy when compared to different baselines,
across different categories, including clothing.

In \cite{AAAIW1715069}, Zhao et. al. try to infer complementary relationship between fashion items
based on the title description. The proposed Siamese Convolutional Neural Network architecture
performs better than other approaches that utilize text-only features, while it requires
minimum feature engineering.

The work of Trofimov, Zahang and Zhao focus mainly on providing static recommendations. There, datasets
that describe the complementary relationship between two items (i.e. \textit{Amazon buy-together}) were used for evaluation.
The domain of personalized complementary recommendation, which takes the user's past history into consideration, is
therefore yet to be fully explored.

Session-based approaches that follow a Recurrent Neural Networks architecture \cite{HidasiKBT15, Hidasi:2018:RNN:3269206.3271761,
Li:2017:NAS:3132847.3132926, Tan:2016:IRN:2988450.2988452}, are found to outperform traditional collaborative filtering
and matrix factorization approaches in domains, where only short sessions are available - such as the e-commerce domain.
More recently, in \cite{Liu:2018:SSA:3219819.3219950} Liu et. al, proposed \textit{STAMP}, a novel Short-Term
Attention Priority Model for Session-based Recommendation that performs in pair of other state-of-the-art approaches
in the offline experiments.

\textit{STAMP}, as well as other session-based approaches, are widely studied in the domain of personalized
recommender systems, due to the nature that a specific customer can be represented by their interaction histories.
In this work, we take the \textit{STAMP} model as the backbone and propose several improvements to forge a variant
that better fits our problem definition and dataset.

\section{Problem Definition}
\label{section:problem-statement}

We define the problem of complementary item recommendations in the Product Display Page as follows:
Given a user $U$ at a given moment of time $t$, it reaches the Product page of the item $x_{t} \in I$ where $I$,
represents the set of possible items.
Before reaching $x_{t}$, the customer interacted with a series of items, denoted as $x_{h}$.
We want to train a recommender $R$ that receives $x_{h}$ and $x_{t}$ as input and returns a sorted list of
$k$ complementary items [$y_0, y_1,... y_{k-1}$] of $x_{t}$ for user $U$.
The recommended list of items will be shown on the Product Display Page (PDP) in an independent carousel.

We propose a definition of complementary items known to work well as a proxy of the worn-together concept: two items
are complementary if they can be worn together.
Specifically, two items $x_i$ and $x_j$ are considered to be complementary if they belong to different nodes on a
category hierarchy\footnote{
The category hierarchy and the negative category pairs encode core business logic of Zalando and are therefore
proprietary and private information.}.
In addition, the categories of items $x_i$ and $x_j$ must not fall into a list of negative category pairs curated by
an in-house Fashion Librarian.
The example shown in Figure \ref{fig:screenshot} shows how a sports shoe, shorts, or jeans
are considered complementary of a t-shirt.
However, two items belonging to the categories of "sports shoes" and "sandals" are not considered as
complementary as this category pair appears in the negative category list.
The proposed approach is agnostic to the definition of the category hierarchy and the negative category pairs.
\section{The STAMP Model}
The \textit{STAMP} model is a neural session-based recommender that takes a sequence of items and a base item as input, and
makes predictions for the very next item in the sequence among a set of candidates $C$.
It creates feature embeddings for all items.
The input sequence is then represented as the attention result of the embeddings of all its member items, denoted by
$x_s$.
Together with the base item embedding $x_t$, the relevance score of a candidate $c_i \in C$, denoted as $x_{c_i}$, is
computed as a trilinear combination.

\begin{equation}
    \begin{aligned}
        & h_s = tanh(W_s^\intercal{x_s} + b_s) \\
        & h_t = tanh(W_t^\intercal{x_t} + b_t) \\
        & z_i =\, <h_s, h_t, x_i> \\
        \label{stamp_score}
    \end{aligned}
\end{equation}
Where $<a, b, c> = \sum_{i=1}^d {a_{i}b_{i}c_{i}}$ and $d$ refers to the dimension of the embeddings. The recommendations
for the given input sequence and the base item is provided by ranking the candidates with their
scores.

\section{Proposed Method}
\label{section:proposed-method}

\textit{CF-c} is known to perform poorly
in terms to attributed purchases. The ratio between the items that are shown in the carousel,
and the items purchased from that same carousel is low. In this section, we describe first how we have
built a training dataset  that accurately models the desired task. We then describe our proposed algorithm.

\subsection{Sampling}
\label{section:sampling}

A common way to compose a complementary item dataset is using the customers' response to the baseline
recommender, and training the model to maximize the accuracy of the next-click prediction.
This approach, though being intuitive and straight forward, introduces the following two drawbacks.
Firstly, due to the poor performance of the baseline algorithm, the generated next-click dataset is not capable of
capturing the common user behavioral patterns.
Secondly, since the dataset encodes mostly the characteristics of the baseline algorithm, it is likely to have a
natural bias towards the baseline.
Therefore, we propose to train and evaluate the models with in our new complementary
item-pair dataset, sampled directly from the general behavior of users in the past.

In order to create a new dataset, we collect a set of user interactions that took place in the past, denoted as $U =
\{u_1, u_2, \cdots , u_m\}$, where $u_i$ represents actions of user $i$ sorted by their timestamps.
Each user action sequence $u_i$ contains user actions $a_1, a_2, \cdots , a_n$, each user action $a_j$
consists of two pieces of information: the actual item id, $x_j$, and the type of interactions $t_j$.
In this study we consider only the most recent $K_c$ click events that happen in the past 9 days and most
recent $K_o$ purchase events that happen in the past 90 days. The sampling strategy works as follows:

\begin{enumerate}
    \item For each user action sequence $u_i$, iterate through each click action $a_j$ and consider actions that
    take place within an hour after $a_j$ as potential targets, denoted as $C_{a_j}$.
    \item User actions in $C_{a_j}$ are filtered out if they do not satisfy the definition of Complementary
    defined by our hierarchy, w.r.t. $x_j$, resulting in a subset $C_{a_j}^-$.
    \item We select target events $T_{a_j} = \{e_{c_1}, e_{c_2}, \cdots , e_{c_o}\}$ from $C_{a_j}^-$ where
    items $x_{c_1}, x_{c_2}, \cdots , x_{c_o}$ are bought in the next 24 hours by this user. These target items are
    used to compose $o$ sequences that share the same item click, purchase history, and the base item $x_j$.
    \item Finally, to rule out potential noisy signals that occur rarely in the dataset, we introduce
    an aditional condition. The target $x_{c_i}$ and the base item $x_j$ have to occur often enough in the overall user history $D$.
    Specifically, we compose a co-occurrence matrix, and we specify that the target item has to be one of the 200
    most frequent items that co-occur with $x_j$.
\end{enumerate}

The cross-sell sequences selected from the users' interaction histories are categorized into training and test set.
The timestamp of a sequence is determined by the time when the base item $x_j$ was clicked. We choose a time-based
division of training and test set, where all sequences from the last day of the sampled data are considered part
of the test set, while the previous sequences are considered part of the training set.

\subsection{Model}

The proposed Complementary Item Recommendations model, \textit{ZSF-c}, includes improvements of the well known
\textit{STAMP}item recommendations model \cite{Liu:2018:SSA:3219819.3219950}, to enhance its performance in the
problem domain.
The main improvements are described as follows.

\subsubsection{Order Events}
A major difference between the problem described in general sequence-based recommenders and our complementary items
recommendation problem is the utilization of the purchase events.
Items purchased by users are intuitively more representative for their long-term tastes and preferences.
Instead of mixing the ordered items together with other clicked items, we separate the purchased items and summarize
them into an independent representation to have separated short term
(the latest 15 number of view events) and
long term (last 5 purchases in the past 3 months) preference of the user.
The representation of orders, denoted as $h_o$ is then combined with the representations of clicked items $h_s$, the
base item representation $h_t$ and the candidate embedding $x_{c_i}$ to calculate the score $z_{c_i}$ between the
user and the candidate item $c_i$.

\begin{equation}
    z_{c_i} = (h_o + h_s + h_t)^\intercal{x_{c_i}}
    \label{sumup_combination}
\end{equation}
Note that instead of using element-wise multiplication to combine the user representations, we sum them,
as we empirically proved that it performs better on our dataset.
By using summation, the final score of an item $i$ represents a combination of how close the item is to the
anchor item $h_t$, the short term history given by views $h_s$ and the long term history given by orders $h_o$.

\subsubsection{Fusing Item Embeddings with Metadata Embeddings}

Unlike STAMP, which uses a single randomly initialized feature vector to represent an item,
the item embeddings in \textit{ZSF-c} come from several information sources.
Categorical features coming from the metadata of an item such as the category and the event type are used to
produce the feature representation of this item.
For an item $i$, denoting its base embedding, category embedding and event type embedding as $m_i$, $g_i$ and $t_i$,
the feature representation $x_i$ is computed through a fusion method.

\begin{equation}
    \begin{aligned}
        & d_i = Concat([m_i,\,m_i \odot g_i,\,g_i]) \\
        & x_i = elu(W_1^\intercal{elu}(W_2^\intercal{d_i} + b2) + b1) \odot (1 + t_i)
        \label{item_embedding_fusion}
    \end{aligned}
\end{equation}

Where $W1$, $W2$, $b1$, $b2$ are learned parameters and are applied to the feature embeddings of all items.

The base embedding $m_i$ can be either randomly initialized, or initialized with pre-trained embeddings.
During training these embeddings are updated together with other model parameters to better accommodate the behavioral
information from the sequences.
The non-linear activation elu \cite{DBLP:journals/corr/ClevertUH15} is selected according to the model performance on
the validation set.
Fusing the categorical features into the item embedding can not only make the model generalize better for new items
or rare items, but also increase the prediction accuracy by better capturing the relationships between different
categories.

\subsubsection{Using Image Features to Initialize the Item Embeddings}

A set of pre-trained fashion-specific image embeddings, denoted \textit{fdna} \cite{KDDFashionDNA}
are used to initialize the base embeddings of all items.
By using the fashion-specific image embedding as an initialization
it preserves several desired characteristics of a fashion product,
such as colors, patterns and shapes.
This information is crucial for the recommender to select complementary products that match the base item.

\subsection{Model Training}

The model is trained on a daily basis to accommodate newly added products and the ever-changing user behavior.
In order to shorten the training time, \textit{Adam} \cite{DBLP:journals/corr/KingmaB14} optimizer is used to
replace the stochastic gradient descent (SGD) algorithm employed in \cite{Liu:2018:SSA:3219819.3219950}.

Even like this, the large number of candidates in our dataset still results in lengthy
training process. However, the maximum training time should remain below 6 hours to avoid having an
stale model in production that can no longer provide up-to-date information to the customers.
To shorten the required training time, a much smaller (typically 2048) randomly selected items are used as negative
samples, and a \textit{softmax} operation is applied to the target and the negative samples to approximate
the global \textit{softmax}.
With this approach it takes around 4 hours to train on the whole dataset for 5 epochs.


While the model trained with the sampled dataset described in section \ref{section:sampling} rarely
produces recommendations within the same category, a filter is added to ensure customers will
only ever see recommendations that adhere to our definition of complementary fashion recommendations.

\subsection{Serving}

We use \textit{TensorFlow-Serving} \cite{Abadi:2016:TSL:3026877.3026899} to load and serve the model in the online test.
Everyday the model training pipeline is triggered once to consume the newly sampled catalog sequences generated from
the latest user interactions, and a new model is trained based on that. The model is automatically downloaded
by our system and passed to the \textit{TensorFlow-Serving} module once the training succeeds.


After retrieving in real-time the most recent user interactions (maintained in a common backend system),
the model is able to select the most suitable 80 out of 120 thousand products to fulfil 500 user requests per second
within 20ms at p99 with two medium-sized CPU instances.

\section{Experiments and Analysis}
\label{section:experiments}


As described in section \ref{section:sampling}, the training and test sets are selected from a consecutive 9 days of
interaction history for all customers who interacted with the Zalando platform during this period of time.
The training set consists of $5073130$ cross-sell examples from $1195512$ users with an average length of $13.08$.
The test set contains $420310$ cross-sell sequences from $131113$ users with an average length of $13.16$.
A validation set, containing $5$\% randomly selected sequences from the training set, is used for hyperparameter
tuning.
The chosen hyperparameters are shown in Table \ref{table:hyperparameters}.


\begin{table}[htb]
    \begin{tabular}{cccccl}
        \hline
        Name && value \\
        \hline
        Learning Rate && 5e-4 \\
        Feature Dimension && 128 \\
        Number of Candidates to rank && 120000 \\
        Number of Epochs && 5 \\
        Activator && elu \\
        Number of Categories && 1400 \\
        Feature Initialization && Xavier \\
        \hline
    \end{tabular}
    \caption{The hyperparameters are selected based on the Order Recall@5 metric in the validation set.}
    \label{table:hyperparameters}
    \vspace{-0.5cm}
\end{table}

\subsection{Offline Results}

We used the test set to evaluate the performance of the proposed model and the \textit{CF-c} baseline according to two main metrics:

\begin{itemize}
    \item \textbf{Recall@}: The percentage of times the clicked item is within the top-k of the recommended items.
    \item \textbf{Order Recall@K}: The percentage of times the clicked item is within the top-k of the recommended items,
    but only for the cases where the clicked item is ordered after the click (within the same day).
\end{itemize}

As the complementary item recommendations carousels display 5 recommendations in its default web layout,
we report both metrics with $K=5$ in our evaluation.
While, users can still see more items by clicking in the "see more" button, only a small percentage of users request more.

The results displayed in Table \ref{table:offline_results} show that \textit{ZSF-c} outperforms \textit{CF-c}
in terms of Order Recall. This metric is critically aligned with the user goal to not only see, but also
wear (and purchase) complementary items. Moreover, previous experiments have shown that the Order Recall metric offline
and online results are more strongly correlated than click-based metric.

\begin{table}[htb]
    \begin{tabular}{c|c|c}
        \hline
        Approach & Recall@5 & Order Recall@5 \\ \hline
        \textit{ZSF-c} & 0.2645 & 0.2673                  \\ \hline
        \textit{CF-c} & 0.2941 & 0.2469                  \\ \hline
    \end{tabular}
    \caption{Offline evaluation results}
    \label{table:offline_results}
    \vspace{-0.5cm}
\end{table}

\subsection{The Ablation Test for Model Improvements}

In order to determine the effect of different model modifications we proposed
in section \ref{section:proposed-method}, we performed an ablation test to see the amount of model performance
gain when a model improvements is applied.
We denote the performance of the original model proposed
by Liu et. al. on the cross-sell dataset as \textit{STAMP},
and the combination of the original model with each independent improvement $X$ is denoted as \textit{STAMP + X}.
The results are shown in Table \ref{table:ablation_results}.

\begin{table}[htb]
    \begin{tabular}{l|c|cl}
        \hline
        Approach & Recall@5 & Order Recall@5 \\
        \hline
        \textit{STAMP} & 0.2217 & 0.2066 \\
        \textit{STAMP} + Order Events & 0.2414 & 0.2238 \\
        \textit{STAMP} + Category Embedding & 0.2582 & 0.2557 \\
        \textit{STAMP} + Image Embedding & 0.2643 & 0.24 \\
        \hline
    \end{tabular}
    \caption{Performance gain from the original \textit{STAMP}  model when different improvements are applied independently.}
    \label{table:ablation_results}
    \vspace{-0.5cm}
\end{table}


\subsection{Online Results}

While the improvements seen in the offline results in terms of Order Recall (also used to decide the best hyper-parameters),
are promising. It is necessary to perform an online A/B test to determine the real performance
of both models in our production environment.


The online experiment was designed to split users in two separate groups. Each group got assigned either the control
(\textit{CF-c}) or the variant approach (\textit{ZSF-c}). The experiment ran for several weeks in the
Zalando Platform within several countries. In order to avoid user behavior pattern shift between weekdays and weekends,
the test ran in full week cycles.

The A/B test was performed using ExPan \cite{expan}, an open-source library statistical analysis of randomised control trials.
These results showed a relative \textbf{improvement of +6.23\% in terms of Click Through Rate (CTR), as well as a relative
improvement of +3.24\% in terms of the number of products ordered from the carousel.}
These results clearly indicate that \textit{ZSF-c} complementary items
recommendations algorithm outperforms the \textit{CF-c} approach, and customers find the items shown to them more relevant.
We observe both an increase in engagement and in order metrics as desired. We believe that the improvement in performance
is thanks to the ability of our model to exploit both the short and long term preferences of the user, as well as being able
to learn dense representations for the items involved in the training, which helps to understand similarity and
complementary relationships.

\section{Conclusions and Future Work}
\label{section:conclusions}

In this paper we present a study of the complementary fashion item recommendation problem,
carried out in a large fashion e-commerce platform.
We describe how we sample sequences from the general actions of users in the platform
to build a new training and testing dataset.
This approach allows personalized session-based models to learn the actual user preferences in terms of complementary
items, mitigating the biases captured by the exiting online complementary products.
Furthermore, we show how we improved STAMP to fit our proprietary dataset, including modifications
done to satisfy the constrains of training time and serving latency typical of a large live enviroment.
We performed a rigorous experiment that compared \textit{ZSF-c} to \textit{STAMP}, as well \textit{CF-c},
the baseline in production at the time of experimentation.
The experiment results show significant improvements in terms of Order-related metrics both in offline and online
evaluation.
Moreover, the online evaluation results also show increase in engagement with the complementary fashion item
recommendations carousel.

In future work, our team plans to further improve the performance of \textit{ZSF-c} by explicitly optimizing the
models with order-specific information.
We also plan to conduct further UX experiments with both customers and fashion experts to guide the next iteration of
the algorithm.

\section{Acknowledgments}

The authors would like to thank the Zalando Recommendations team, as well as
all other colleagues involved in the process of building and testing the proposed approach,
and writing and reviewing this paper.

    \bibliographystyle{ACM-Reference-Format}
    \bibliography{paper}

\end{document}